\documentclass{emulateapj}
\usepackage{graphics,graphicx}
\usepackage{ulem}
\def\kms{km\,s$^{-1}$}
\def\etal{ et~al.\rm}

\def\mjysr{MJy sr$^{-1}$}

\def\g292{G292.0+1.8}
\def\spitzer{{\it Spitzer}}
\def\herschel{{\it Herschel}}

\def\chan{{\it Chandra}}
\def\msun{$M_{\odot}$}

\tighten
\begin{document}

\title{The Cold Dust Content of the Oxygen-Rich Supernova Remnant G292.0+1.8  }

\author{Parviz Ghavamian\altaffilmark{1},  Brian J. Williams\altaffilmark{2} 
  }

\altaffiltext{1}{Department of Physics, Astronomy and Geosciences,
  Towson University, Towson, MD, 21252; pghavamian@towson.edu}
\altaffiltext{2}{CRESST/USRA and X-ray Astrophysics Laboratory, NASA
  Goddard Space Flight Center, Greenbelt, MD, 20771;
  brian.j.williams@nasa.gov}

\begin{abstract}

We present far-infrared images of the Galactic oxygen-rich supernova remnant (SNR)
\g292, acquired with the PACS and SPIRE instruments of the \herschel\,
     {\it Space Observatory}.  We find that the SNR shell is detected in the PACS blue (100
     $\micron$) band, but not in the red (160 $\micron$) band, broadly consistent with results
     from AKARI observations.  There is no discernible emission from
     \g292\, in SPIRE imagery at 250, 350 and 500 $\micron$.   Comparing the 100 $\micron$
     emission to that observed with \spitzer\, at 24 and 70 $\micron$, we find a very similar appearance
     for \g292\, at all three wavelengths.  The IR emission is dominated by dust from
     non-radiative circumstellar shocks.  In addition, the radiatively shocked O-rich clump known as the 'Spur' on the
     eastern side of \g292\, is clearly detected in the PACS blue
     images, with marginal detection in the red.   Fitting the existing 14-40 $\micron$ IRS spectra of the Spur together
     with photometric measurements from 70 $\micron$ MIPS and 100 $\micron$ PACS photometry,  
     we place an upper limit of $\lesssim$0.04 \msun\, of ejecta dust mass in the Spur, under the most conservative assumption
     that the ejecta dust has a temperature of 15~K.  Modeling the dust continuum
     in the IRS spectra at four positions around the rim, we estimate postshock densities
     ranging from $n_p\,=\,$ 3.5 cm$^{-3}$ to 11 cm$^{-3}$.   The integrated
     spectrum of the entire SNR, dominated by swept up circumstellar dust, can be fit with a two-component dust
     model with a silicate component at 62 K and graphite component at
     40 K for a total dust mass of 0.023 \msun.

\end{abstract}


\section{Introduction}

The Galactic supernova remnant (SNR) \g292\, was producted by an
unrecorded SN that occurred roughly 3000 years ago (Ghavamian et
al. 2005; Winkler \& Long 2006; Winkler et al. 2009).  It is
classified as an oxygen-rich SNR based on the dominance of oxygen
line emission in its optical spectrum.  In [O~III] $\lambda$5007 \AA\, optical images \g292\,
is dominated by a distinct, crescent-shaped structure approximately
1\arcmin\, in size (known as the 'Spur') located on the eastern side
of the SNR (Goss \etal\, 1979).  Modeling of spectra in the optical
(Winkler \& Long 2006) and mid-infrared (Lee \etal\, 2009; Ghavamian \etal\, 2009; 
Ghavamian \etal\, 2012) indicates a wide range of densities ($\sim$0.5-100 cm$^{-3}$) and shock
velocities ($\sim$ 20-200 \kms) throughout the Spur.  A collection of
localized ejecta clumps (fast-moving knots, or FMKs) have also been found in the interior of \g292\,
(Ghavamian et al. 2005; Winkler \& Long 2006).  This contrasts with
the morphology of the SNR in X-ray (Clark \etal\, 1980; Park \etal\,
2002; 2004, 2007; Kamitsukasa \etal\, 2014; Bhalerao \etal\, 2015),
radio (Gaensler \& Wallace 2003) and infrared (Lee \etal\, 2009;
Ghavamian \etal\, 2009, 2012) wavelengths, where the remnant appears as
a slightly elliptical shell with maximum width of 8\arcmin.  The
shell is approximately bisected by a belt of circumstellar material running E-W.  The current
consensus is that the shell is likely the relic of the progenitor
star's equatorial winds (Park \etal\, 2002; 2004; 2007; Gonzalez \& Safi-Harb 2004).

Here we report on far-infrared imaging observations of \g292\ obtained
with the PACS and SPIRE instruments on the {\it Herschel Space Observatory}. 
The PACS images were taken in the 85-130 $\micron$ (blue, centered on
100 $\micron$) and 130-210 $\micron$ (red, centered on 160 $\micron$)
bands.  Emission at these wavelengths traces mostly the cool dust ($\sim$20-30 ~K)
heated by shocks in the circumstellar medium (CSM).  Our observations
also included SPIRE imagery at 250, 350 and 500 $\micron$, meant to
probe the coldest dust ($\lesssim$20~K).   The main results of our observations are 
the successful detection of cold dust in the radiatively shocked supernova ejecta of \g292, as well as the successful detection 
of cold dust in the circumstellar shocks surrounding \g292.

\section{Observations and Data Reduction}

The observations of \g292\, described in this paper were performed during OT2 
of \herschel\, (PID 40583; P. Ghavamian, PI).  They consisted of large scan maps of the
SNR with the PACS instrument in imaging photometer mode on 2013 January 5, and 
included observations in blue (100 $\micron$) and red (160 $\micron$)
channels.  Additional observations of \g292\, were performed with the SPIRE instrument on 2012 June 23
 at 250, 350 and 500 $\micron$.  The PACS observations used a medium speed map with 5 scan legs of length
10$^{\arcmin}$, with cross-scan steps separated by 110$^{\arcsec}$.
The diameter of \g292\, is approximately 8.3$^{\arcmin}$ (14 pc at an assumed
distance of 6 kpc; Gaensler \& Wallace 2003), so the PACS scan map
was constructed to allow sufficient coverage of the entire remnant and nearby
background. The PACS scan map was repeated 80 times for a total
exposure time (on-source) of 12,000 s.  The 1-$\sigma$ extended source
sensitivity of the PACS observations was approximately 1.8 \mjysr\, at
100 $\micron$ and 1.1 \mjysr\, at 160 $\micron$.  The SPIRE images
were acquired with a nominal speed scan map with 8\arcmin\, across and
repeated 20 times for an on-source total exposure time of 1,649 s.

Our PACS data were processed with \herschel\, Interactive Processing
Environment (HIPE) and converted to level 3 mosaics.  The pixel scale
of the mosaics was 1.6$^{\arcsec}$ pixel$^{-1}$ in the blue channel
and 3.2$^{\arcsec}$ pixel$^{-1}$ in the red channel.  The diffraction
limited optics produce a full width at half maximum (FWHM) of
8$^{\arcsec}$ for the PACS PSF at 100 $\micron$ and 12$^{\arcsec}$, at
160 $\micron$. The flux calibration uncertainty for PACS is less than
10\% (Poglitsch et al. 2010) and the expected color corrections for the PACS 100 and 160 $\micron$ channel are
small compared to the calibration errors ($\lesssim$7\% for a blackbody with 20~K\,$\lesssim\,T\,\lesssim$\,50~K). We therefore adopt a 10\% calibration error.  The SPIRE data were also processed with HIPE and
converted to level 2 products having pixel scale of 6$^{\arcsec}$
pixel$^{-1}$, 10$^{\arcsec}$ pixel$^{-1}$ and 14$^{\arcsec}$
pixel$^{-1}$ at 250, 350 and 500 $\micron$, respectively.  The
corresponding PSF FWHM for the three SPIRE bands are 14$^{\arcsec}$ at
250 $\micron$, 25.2$^{\arcsec}$ at 250 $\micron$ and 26.6$^{\arcsec}$
at 250 $\micron$, respectively.

To enable comparison of photometry between the blue channel PACS image and the
existing 70 $\micron$ \spitzer\, MIPS image of \g292\, (Ghavamian \etal\, 2012) we first used the IRAF
task WREGISTER to transform the PACS 100 $\micron$ image to the same
spatial scale and orientation as the MIPS 70 $\micron$ image.  We then
binned the transformed image to the same pixel scale as the latter,
and finally convolved the binned image with the appopriate kernel to match the PSF
of the MIPS 70 $\micron$ image.  We used the IDL-based Convolution
Kernels software CONVIMAGE (Gordon \etal\, 2008) with a kernel
specifically designed to smear the PACS 100 $\micron$ PSF to that of
the MIPS 70 $\micron$ PSF.  To extract surface brightness values from
 \g292\, we utilized the FUNTOOLS package of SAO ({\sf https://www.cfa.harvard.edu/\textasciitilde
  john/funtools/}).

As a supplement to our analysis in this paper, we have combined PACS photometric measurements
from specific regions of interest around \g292\, with spectra extracted from the \spitzer\, IRS spectral imaging datacube (Ghavamian \etal\, 2012).   The basic configuration used for those observations is described as follows.   The IRS spectral observations consisted of mapping 
scans with both 1st and 2nd order Long-Low slits, which together covered the entire SNR in the 14-40 $\micron$ wavelength range.  
There were 560 individual spectra acquired, each at an exposure time of 32s pixel$^{-1}$.     The analysis of that dataset is described 
in detail in Ghavamian \etal\, (2012), and the reader is referred to that paper for more information.  

\section{Imaging Results}

The mosaicked level-3 images of \g292\, are shown side by side in Figure~1.  The morphology of \g292\, at 100 $\micron$ is strikingly similar 
to its appearance in the MIPS 24 $\micron$ and 70 $\micron$ bands.  An ellipsoidal shell, which coincides with the shell of emission 
seen in X-rays, can be seen at 100 $\micron$.
The shell extends nearly three quarters of the way around \g292,
fading to invisibility along the eastern quarter of the SNR.  The lack of far IR
emission on the eastern side of \g292\, is consistent with results
from X-ray analyses, which indicate much lower preshock densities in
that region than elsewhere along the rim (Lee \etal\, 2010).  The equatorial belt, noted
in previous X-ray (Park \etal\, 2004; Park \etal\, 2007, Lee \etal\,
2010)  optical (Ghavamian \etal\, 2005) and infrared (Ghavamian
\etal\, 2009; Lee \etal\, 2009; Ghavamian \etal\, 2012) observations can be seen stretching across 
the middle of \g292\, at 100 $\micron$.  Detailed analysis of the shell at X-ray  (Lee \etal\, 2010) and mid-infrared wavelengths 
(Ghavamian \etal\, 2012) showed an elevated preshock density along the southwestern
edge of \g292, consistent with features seen in the PACS 100
$\micron$ image.  No portion of the shell or
equatorial belt is detected at 160 $\micron$ (Figure~1).  Emission clumps are seen both along the
southwestern edge of \g292\, and immediately outside the shell.    Although
some of the clumpy emission knots along the southwestern edge of
\g292\, at 100 $\micron$ are also present at 160 $\micron$, they have
no X-ray or optical counterparts, thus indicating that their emission
is not shock excited.  They may be molecular cloud structures along the line of
sight unrelated to \g292.  However, the elevated preshock density in this part of
the SNR (Lee \etal\, 2010) may indicate some connection with \g292. 

No emission is detected from \g292\,in any of the three SPIRE bands
(Figure~2).  The clumpy emission knots seen outside the southwest corner
of the remnant in the PACS images are also seen at 160, 250,
350 and 500 $\micron$, consistent with dense, cold clouds heated by
ambient ultraviolet radiation. Again, the clumpy knots may
be connected to a larger structure such as a molecular cloud along the line of sight.  

No discernible emission is detected from unshocked ejecta dust in any of the \herschel\, images.  
This dust component is predicted to exist interior to the reverse shock remnants of core collapse
explosions, and can be heated to temperatures $\sim$30-40~K from photoionizing
radiation emitted by the overlying shocked ejecta (Lee \etal\, 2015).  Emission from cold unshocked ejecta dust 
was detected in \herschel\, PACS observations of Cassiopeia A (Barlow \etal\, 2010).    However, in \g292\, emission 
from the cold ejecta dust falls below the PACS detection limit.  

Finally, the radiative shocks in the Spur are clearly detected at 100 $\micron$ (Figure~1).   The origin of this
emission is discussed in Section~5.  There is no obvious IR emission from the extensive X-ray emitting
(shocked) ejecta.  

\section{Comparison to Optical and X-ray Imagery}

Many of the shock-heated circumstellar structures observed in \g292\, at X-ray wavelengths are also detected
in the PACS 100 $\micron$ band.  The PACS 100 $\micron$ image is shown alongside a deep \chan\, image
(720 ks; Park \etal\, 2007) in Figure~3.  X-ray spectral analyses of
these structures (Park \etal\, 2004, 2007; Lee \etal\, 2010) found all of them
to be of interstellar composition, an indication that their emission arises from shocked  
circumstellar gas.  The features in common between the two bands are marked by arrows and numbered 
 C1 through C9 in Figure~3 ('C' indicating 'circumstellar').  The outer shock wave (C1, C2, C5,
C6, C10) is clearly detected in both the far IR and soft X-rays, as is
the equatorial belt (C8) and associated structures (C7, C9).  The
density of the equatorial belt ($n_H\,\sim$ 5 cm$^{-3}$, Park \etal\,
2004) is elevated relative to the outer blast wave shell
($n_H\,\sim$0.8-1.2 cm$^{-3}$; Lee \etal\, 2010), with correspondingly
lower shock speeds in the belt ($\sim$500 \kms; Ghavamian \etal\, 2005)
compared to the shell ($\sim$1500-2000 \kms; Lee \etal\, 2010).  Under
these conditions, shocks in the belt have begun to form recombination
zones, consistent with the faint optical [O~III] and
H$\alpha$ emission detected from structures C7, C8 and C9 (Ghavamian \etal\,
2005).  At least two compact knots of emission near the southwestern
edge of \g292\, (C3, C4) are also detected in both bands.  The \chan\,
spectra of these knots shows that they are of cosmic (sub solar)
composition (Park 2010: private communication). It is unclear whether
they are simply interstellar clouds, or whether they are part of the
progenitor star's wind.  If the latter, then that would indicate a
highly clumpy stellar wind structure.

None of the shocked ejecta seen at X-ray wavelengths exhibit any
obvious counterpart in the PACS or SPIRE images, indicating that the
dust within the X-ray ejecta is either too cold to emit significantly
in those bands, or that it does emit in those bands but falls below the detection limit of  \herschel\,
in the allotted exposure time.  The dust in the X-ray emitting ejecta is expected to be cold due to the low plasma density
(Dwek, Foster \& Vancura 1996; Borkowski \etal\, 2006; Williams \etal\, 2011).  In Cassiopeia A, on the other hand, dust emission has been detected by \spitzer\, from the X-ray
emitting ejecta (Smith \etal\, 2009).   The difference is partly due to the fact that \g292\, is nearly 5 times older than Cassiopeia A, so that its ejecta
have expanded to lower density than those of Cassiopeia A.  

Although the Spur is detected at 100 $\micron$, none of the fainter optically emitting ejecta knots (such as the fast-moving knots, or FMKs) are detected.
The difference is likely due to the higher density of the radiative shocks in the Spur compared to the FMKs,
as found by Winkler \& Long (2006) and Ghavamian \etal\, (2009).   The higher gas density results in a higher temperature for the dust there,
and hence greater emissivity in the PACS 100 $\micron$ band compared to the FMKs.  

A magnified view of the region around
the Spur (marked by the dotted square in Figure~3) is shown in
Figure~4 in optical [O~III] (Winkler \etal\, 2006), side by side with
the PACS 100 $\micron$ and PACS 160 $\micron$ images.  Radiatively
shocked O-rich ejecta are labeled E1-E6 ('E' referring to 'ejecta').  There is striking
correlation between the brightest [O~III] features at the top of the
Spur (E1 and E2) and the corresponding features at 100 $\micron$.
Faint emission from these features can also be discerned in the 160
$\micron$ image.  Other, fainter [O~III] clumps farther down the Spur
(E3, E5 and E6) are clearly seen at 100 $\micron$, though not at 160
$\micron$ due to a combination of their faintness and lower spatial
resolution at 160 $\micron$.

In addition to [O~III] $\lambda$5007 \AA\, emission, the oxygen-rich radiative shocks in the Spur can be expected to 
produce [O~III] 88 $\micron$ emission.    Most of the oxygen line emission is expected to arise from
a region of nearly constant electron temperature ($\sim$10$^5$~K) behind the radiative shocks (Itoh
1981, 1988; Borkowski \& Shull 1990).  The one oxygen-rich SNR with a spectroscopically measured [O~III] 88 $\micron$ flux 
is Cas A, where the observed ratio, corrected for interstellar reddening toward Cas A ($A_V\,\approx\,$ 5, Hurford \& Fesen 1996) , 
is $F_{[O~III]}(88\, \micron)/F_{[O~III]}(5007 \AA)\,\sim\,$0.1-0.14 
(Docenko \& Sunyaev 2010).  If we assume an intrinsically similar ratio for \g292, then applying the appropriate interstellar reddening for \g292\,
($A_V\,\approx\,$ 2, Winkler \& Long 2006), the predicted ratio of 88 $\micron$ to 5007 \AA\, emission for \g292\, should only be $\sim$\,0.02
(due to the lower impact of reddening on the 5007 \AA\, line in \g292).
Scaling this ratio now to the [O~III] $\lambda$5007 \AA\, surface brightness of clumps E1 and E2 in \g292\, (Winkler \etal\, 2006) results in a predicted
[O~III] 88 $\micron$ surface brightness that is at least an order of magnitude lower than observed in the PACS 100 $\micron$ image.     Therefore,
we do not expect a substantial contribution from [O~III] 88 $\micron$ to the emission observed from the Spur at 100 $\micron$.

Finally, we considered the contribution of [O~I] 63 $\micron$ emission to the 100 $\micron$ image.  The observed ratio  
$F_{[O~I]}(63\, \micron)/F_{[O~III]}(5007 \AA)$ after correction of interstellar reddening toward Cas A is approximately 0.07-0.1  (Docenko \etal\, 2010), 
so again under the assumption that \g292\, exhibits a similar ratio, the [O~I] 63 $\micron$ surface brightnesses of clumps E1 and E2 are predicted to be at 
least an order of magnitude lower than their observed surface brightnesses in the 100 $\micron$ image.  Further reducing the likelihood of [O~I] contamination 
is the transmission of the PACS  channel at 63 $\micron$, which is only $\sim$ 0.5\%\,(Poglitsch \etal\, 2010).  Therefore, we do not expect significant contribution 
from [O~I] in the PACS 100 $\micron$ image either.  Of course, we cannot for certain know the contributions of [O~I] and [O~III] lines to the far-infrared emission
in the PACS image without a far infrared spectrum of the the ejecta.  However, it appears reasonable to assume that the PACS
blue band emission from the Spur arises from dust continuum within the ejecta, most likely formed during the SN explosion and now heated in the
radiative shocks.

\section{Analysis}  

The background-subtracted photometric global fluxes from \g292\, at 24, 70 and 100 $\micron$ are shown in Figure~6.  The fluxes at 24 and 70
$\micron$ are taken from Ghavamian \etal\, (2012), while the PACS 100 $\micron$ flux is taken from the PACS image convolved to MIPS 70
$\micron$ resolution (see Table 2).  An upper limit on the flux at 160 $\micron$ as measured from the PACS red channel is marked by the inverted triangle.

\subsection{Dust Models of the Oxygen-Rich Ejecta in the Spur}

The cold dust present in the oxygen-rich Spur is likely associated with radiatively shocked ejecta (Ghavamian \etal\, 2009, 2012), where
the heating source of the dust would not be collisional heating by energetic particles, but radiative heating from the slow, radiative shocks.
Physically modeling such emission is beyond the scope of this paper.  However, we can still obtain useful information about the
total mass and temperature of the radiating dust, useful for constraining how much dust can be formed in core-collapse supernovae. 

We extracted the IRS spectrum from the Spur, along with the MIPS flux at
70 $\micron$ and the PACS fluxes at 100 and 160 $\micron$ (with the 160
$\micron$ data point being only an upper limit).   Both the IRS spectral extraction and PACS flux measurements were obtained
from a rectangular region centered on the Spur, using a nearby background region located 
just outside the remnant.  The SED of the Spur is shown in Figure~7 and the extracted fluxes are listed in Table 2.    

We note that the spectra show a clear sign of silicate dust, the ``shoulder'' in the spectrum seen at $\sim 18$
$\micron$. We therefore assumed a pure generic silicate dust, and fit with a two-temperature model. 
While it is true that a pure silicate grain model would not account for the 
possibility of a small amount of CSM dust projected along the line of sight to the Spur, even this dust would consist 
mostly of silicate grains (Weingartner \& Draine 2001).  As mentioned above, our aim is not to provide a detailed dust census, but to 
provide an upper limit on the total amount of dust present in the Spur. In this model we assumed a single grain size of
0.05 $\micron$, which is unphysical, but Temim \& Dwek (2013) point out
that such an assumption only serves to make the derived dust masses
upper limits to the actual value, which again, is the parameter we seek for the Spur. We obtained the fit shown in Figure~7 with 
the combination of a hot component, with $T_{hot} = 114$ K and $M_{hot}$ = $5 \times 10^{-6}$
\msun, and a cool component, with $T_{cool} = 52$ K and $M_{cool} = 1.2 \times 10^{-3}$ \msun.  

Finally, we estimated an upper limit on the dust mass under the most extreme assumption that 
a third, even colder, undetectable component were present. For this, we fixed the hot and cold components at their
fitted values, but added  a hypothetical dust component with an extreme temperature of T $=$ 15 K. We allowed the mass of
this component to be as high as possible, without violating the measured upper limit on 160 $\micron$.  Under this assumption,
we found that the mass of this cold component was limited  to $< 0.04$ \msun. 

\subsection{The Global Dust Spectrum}

In the next part of our analysis, we considered the total amount of IR flux, integrated from the
entire remnant. We constructed an SED using fluxes at 24, 70 from the estimate of Ghavamian \etal\, (2012),
as well as the 100, and 160 $\micron$ estimates (shown in Figure~6).   
The emission line contribution is negligible, so that virtually all of the
emission is produced by dust. As we showed above, this dust
arises from different components (both ejecta dust and CSM dust), and
even within the CSM dust, there are a variety of different physical
conditions. It would be impossible to construct a unique physical
model for all these different components, so we simply fit the SED in
a similar fashion to that done for the ejecta dust in the Spur, with
one modification: since the vast majority of the emission in the
remnant is dominated by the bright CSM dust emission, we used a
two-component model consisting of a silicate component and a graphite
component, with the mass ratio between the two fixed at 2.5
(Weingartner \& Draine 2001).

We find that the global spectrum of \g292\, can be fit well with a
two-temperature dust model with T$_{silicate}$ = 60~K, T$_{graphite}$
= 39~K (green and red dashed curves in Figure~6), with a total dust mass M$_{dust}$ = 0.023
\msun.  Again, this includes all sources of dust: CSM, equatorial
belt, and supernova ejecta.  By comparison, Lee \etal\, (2009) obtained
similar results using a carbonaceous/silicate grain mix as prescribed
by Draine \etal\, (2003) and applying a two-temperature dust model to
the AKARI global spectrum.  They obtained a total dust mass of
4.5$\times$10$^{-6}$ \msun\, for a warm component (T\,=\,103~K) and
$\lesssim$0.048 $M_{\sun}$ for a cold component (T\,$\sim$47~K).  

The estimated mass of swept-up gas for \g292\, is 15-40 \msun\, (Lee \etal\, 2010), which combined 
with our dust mass estimate gives a dust to gas ratio of (0.6-1.5)$\times$10$^{-3}$.  This is
significantly lower than the average dust-to-gas ratio of 6.2$\times$10$^{-3}$ (Zubko \etal\, 2004) for the interstellar medium,
indicating a significant amount of dust destruction in the circumstellar shocks around \g292. This is not unusual; for example,
in a study of IR emission from Kepler's SNR with \spitzer, Blair et al. (2007) found that 78\% of the pre-existing CSM dust has 
been destroyed by the passage of the shock wave.

\subsection{Dust in the Bright Circumstellar Ring}

In the last part of our analysis we focused on modeling IRS spectra from four localized portions of the bright
circumstellar ring (marked on Figure~5).  These regions were chosen to sample the
morphological variations observed in the \spitzer\, images.   The shocks in these regions 
are some of the most clearly defined structures discernible in the IR.  
The regions include some of the
faintest emission from the outer blast wave along the northern edge
(Region 1), as well as circumstellar shocks with enhanced emission
along the western rim (Regions 3 and 4).  The emission in Region 2 is
intermediate in surface brightness between the northern and western
regions. In each case we subtracted a local background from a region
of the same size, located immediately outside the remnant.  One
motivation for choosing localized regions, rather than a global
average, was to take into account variability of the background
emission around \g292.  As noted by Ghavamian \etal\ (2012), the
background around \g292\, exhibits a positive surface brightness gradient running from E to W, correlated with the relatively higher
surface brightness of the circumstellar shocks on the eastern side.
We extracted IRS spectra for those regions from our IRS data cube and then subtracted an
off-source background, as marked in Figure~5.  The resulting background-subtracted IRS
spectra are shown in Figure~8.

To model the dust spectra, we used a procedure identical to that in our
previous work; see Borkowski et al. (2006) and Williams \etal\, (2012) for a full
description. The heating of grains in the post-shock plasma by collisions
with hot ions and electrons is modeled,  taking grain size distributions
for both silicate and graphite grains in the Milky Way in appropriate
proportions (Weingartner \& Draine 2001).  The shock model calculates both the
heating and subsequent destruction of grains due to sputtering,
and predicts an output IR spectrum. The primary inputs are gas
temperature, gas density, and ionization age of the plasma (defined as
$\tau_{p} = \int^{t}_{0} n_{p} dt$, where $n_{p}$ is the post-shock
proton density). We fixed the electron temperature and ionization age in
each region to those derived in Lee et al. (2010) (though note that the
variations in these parameters have fairly small effect on the resulting dust spectrum). The proton
temperature is assumed to be 5 keV for all regions, though this has an
almost completely negligible effect on the fit (Williams et
al. 2011). The post-shock density, which has by far the largest
effect, was then left as a free parameter to tune to obtain the best
fit to the spectrum. We assumed cosmic abundances, so that the
post-shock electron density is 20\% higher than that of the protons.

The post-shock densities predicted by our fits are listed, with uncertainties, in
Table 1, and the resulting fits to the IRS spectra are shown in Figure 8. The regions we 
selected for analysis differ from spatial regions examined by Lee \etal\, (2010);
see Figure 4 of their paper. Our densities are a few times higher
than those obtained by Lee \etal\, (2010) from the X-ray fits.  However,
this is to be expected since their regions sample the more
rarefied emission of the outermost blast wave. Additionally, our
regions likely are a superposition of emission from both components
(i.e., circumstellar belt and more diffuse ambient medium just behind the blast wave
shock). Given that G292.0+1.8 is believed to be expanding into an
$r^{-2}$ stellar wind, we would expect the densities interior to the
blast wave emission to be higher. The densities we
derive for the belt are quite similar to those derived for that structure
by Park et al. (2007).

\section{Summary}

We have presented PACS and SPIRE imaging observations of \g292\,
obtained with \herschel.  We have combined these observations with
\spitzer\, datacube spectra (14-40 $\micron$) to estimate the quantity
of cold and warm dust in \g292.  We have taken advantage of the superior spatial
resolution of \herschel\ PACS to identify regions of localized dust
emission within the radiatively shocked O-rich ejecta in \g292.  This
allows us to identify the presence of cold dust specifically associated with the 
O-rich radiative shocks. 

Most of the cold dust in the remnant is morphologically associated with the
prominent shell of circumstellar material identified in X-ray
images. Thus, most of this dust is pre-existing, rather than
newly-formed, and is being heated and destroyed by the forward shock
into this shell. We use the dust spectra from a few regions along the shell to derive the gas densities there, finding them to be
$\sim 4-10$ cm$^{-3}$, a few times higher than the densities of the
emission from the elliptically shaped forward shock.  This is consistent with the fact
that the shell is substantially brighter in both X-rays and IR than the outermost material.

We find only a small amount of dust, about a thousandth of a solar
mass, associated with the O-rich ejecta identified in optical images
of the remnant. This may not account for all the ejecta dust present,
but even if we consider {\it all} of the IR emission from the entire
remnant, we obtain a total dust mass of only 0.023 \msun.

\g292\, is only the third oxygen-rich supernova remnant, aside from 1E0102$-$72.3 (Sandstrom \etal\, 2009; Rho \etal\, 2009)
and Cassiopeia A (Rho \etal\, 2008; Smith \etal\, 2009), having supernova ejecta dust detected at 100 $\micron$ with \herschel.
Unlike the well known O-rich supernova remnant Cassiopeia A, where IR dust continuum
has been detected by \herschel\, in all three components of the ejecta (radiative
shocks, non-radiative (X-ray emitting) shocks and cold unshocked ejecta, Ennis \etal\, 2006; Smith \etal\, 2009), dust continuum
from the ejecta in the much older SNR \g292\, is only detected from the densest
radiative shocks in the Spur.  This is consistent with the much greater age, and hence lower ejecta density
of \g292\, compared to Cassiopeia A.  

The authors wish to thank John Raymond for valuable discussions regarding the
interpretation of the {\it Herschel} data, as well as the anonymous referee for patient reading of the manuscript and very helpful feedback.  This 
work is based in part on observations made with {\it Herschel}, a European Space Agency Cornerstone Mission with significant participation by 
NASA.   Support for this work, part of the NASA Herschel Science Center Theoretical Research/Laboratory Astrophysics Program, 
was provided by NASA through a contract issued by the Jet Propulsion Laboratory, California Institute of Technology under a contract with NASA.

\clearpage

\begin{deluxetable}{ccc}
\tablecaption{Derived Densities From Modeling of IRS Spectra}
\tablehead{
\colhead{Region} & $n_{p}$ (cm$^{-3}$)  }
\startdata
\tablewidth{30pt}
\\
1 & 3.5$^{\,3.9}_{\,3.2}$\\
2 & 4.2$^{\,4.5}_{\,3.9}$\\
3 & 11.0$^{\,11.3}_{\,10.6}$\\
4 & 10.3$^{\,10.5}_{\,10.1}$\\
\tablewidth{100pt}
\enddata \tablecomments{Post-shock proton densities for the four spectra fit
in Figure 8. Uncertainties are 90\% confidence limits from
  $\chi^{2}$ fits to IRS spectra.}
\label{densitytable}
\end{deluxetable}

`````\begin{deluxetable}{ccccc}
\tablecaption{Fluxes Utilized in Model Fits in Figures 6 and 8}
\tablehead{
\colhead{Region} & F(24) (Jy) &  F(70) (Jy)  &  F(100) (Jy)   &   F(160) (Jy)  }

\startdata

Global SED & 9.8$\pm$0.98$^a$ 	& 26.4$\pm$4.2$^a$ 	&  20.2$\pm$3.0	&  $\leq$8.5 \\
Spur	 & \nodata $^b$ 	&	\nodata $^b$ 	&	1.26$\pm$0.2	&	0.57$\pm$0.07

\enddata \tablecomments{ a) Fluxes as reported in Ghavamian \etal\, (2012).   b) Models are fit to IRS spectra at these wavelengths}
\label{densitytable}
\end{deluxetable}


\begin{figure}[ht] 
 \centering
\includegraphics[width=7in]{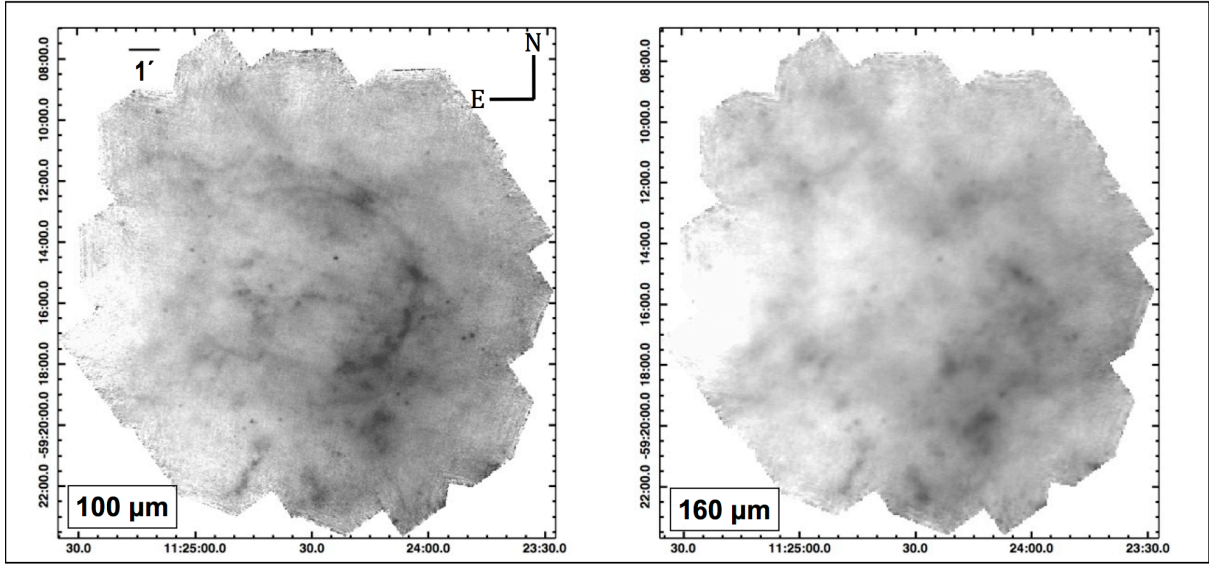}  
   \caption{ PACS blue band (left) and red band (right) images of
     \g292.  The outer shock can be seen on the western side as a
     partially complete elliptical shell on the western edge of the
     SNR.  }
   \label{fig:fig1}
\end{figure}

\begin{figure}[ht] 
\centering
\includegraphics[width=7in]{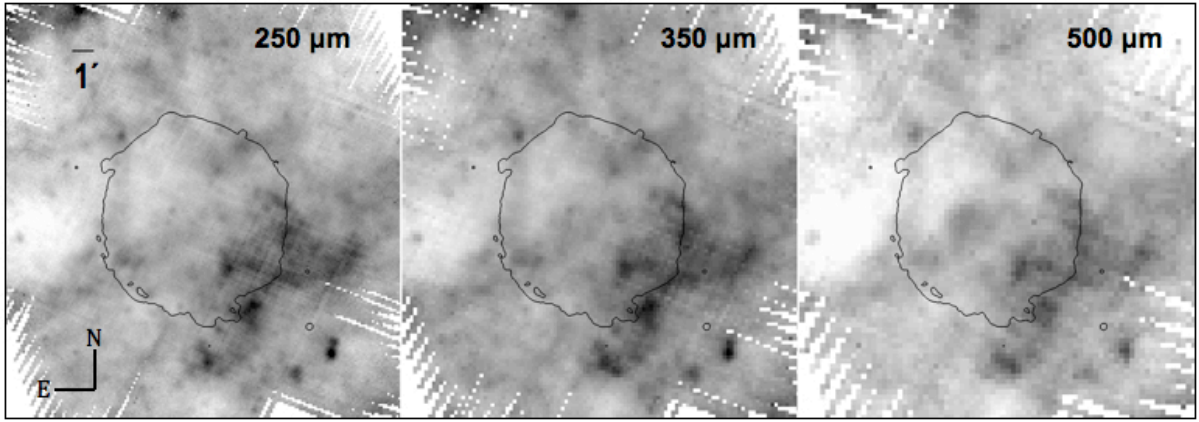}  
\vskip -1.0in
   \caption{ SPIRE images of \g292, shown at 250, 350 and 500
     $\micron$ with the outermost soft X-ray contour (0.3-1.0 keV; Park \etal\, 2007) marked.
     The images are dominated by unrelated background/foreground
     emission, with no discernible contribution from the SNR.  }
   \label{fig:fig2}
\end{figure}

\begin{figure}[ht] 
\centering
\includegraphics[width=6in]{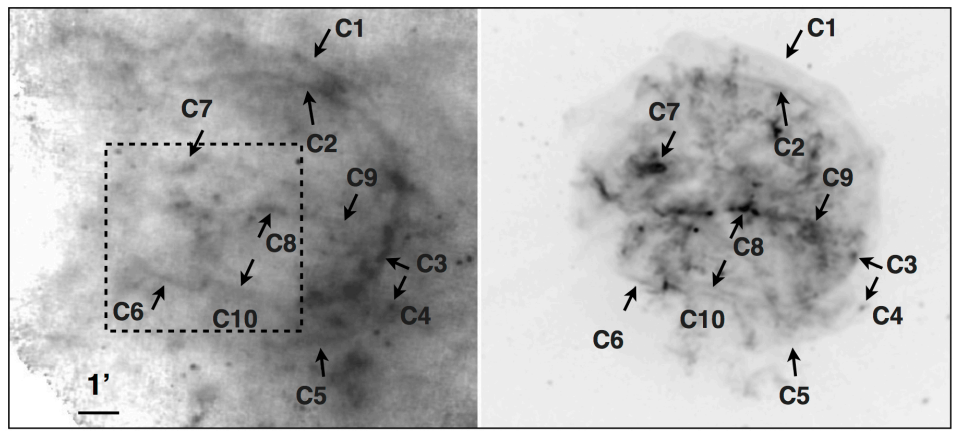}  
   \caption{ Comparison of the 100 $\micron$ PACS and the deep X-ray \chan\,
   images (Park \etal\, 2007) of \g292.  Prominent features discernible in both bands are 
   marked C1-C10.   The dashed square indicates the region used for the closeup of the oxygen-rich Spur in Figure~4.}
   \label{fig:fig3}
\end{figure}

\begin{figure}
\centering
\includegraphics[width=7in]{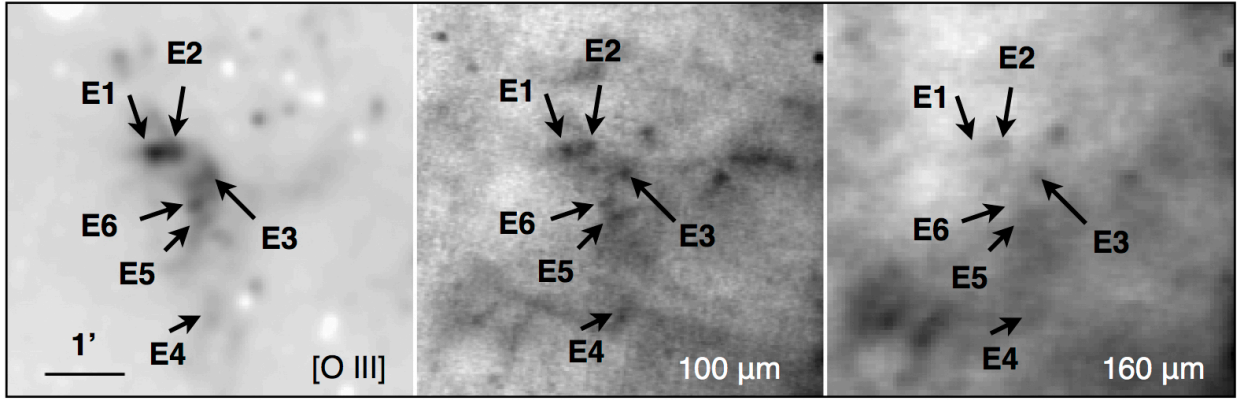}  
\vskip -2.4in
   \caption{ Closeup of the Spur, a dense clump of radiatively shocked
     ejecta, shown in [O III] (left; Winkler \etal\, 2006), at 100 $\micron$
     (center) and 160 $\micron$ (right).  Most of the Spur is
     discernible at 100 $\micron$.  Prominent features discernible in both [O III] and
     100 $\micron$ are marked E1-E6. 
   }
   \label{fig:fig4}
\end{figure}

\begin{figure}
\centering
\includegraphics[width=7in]{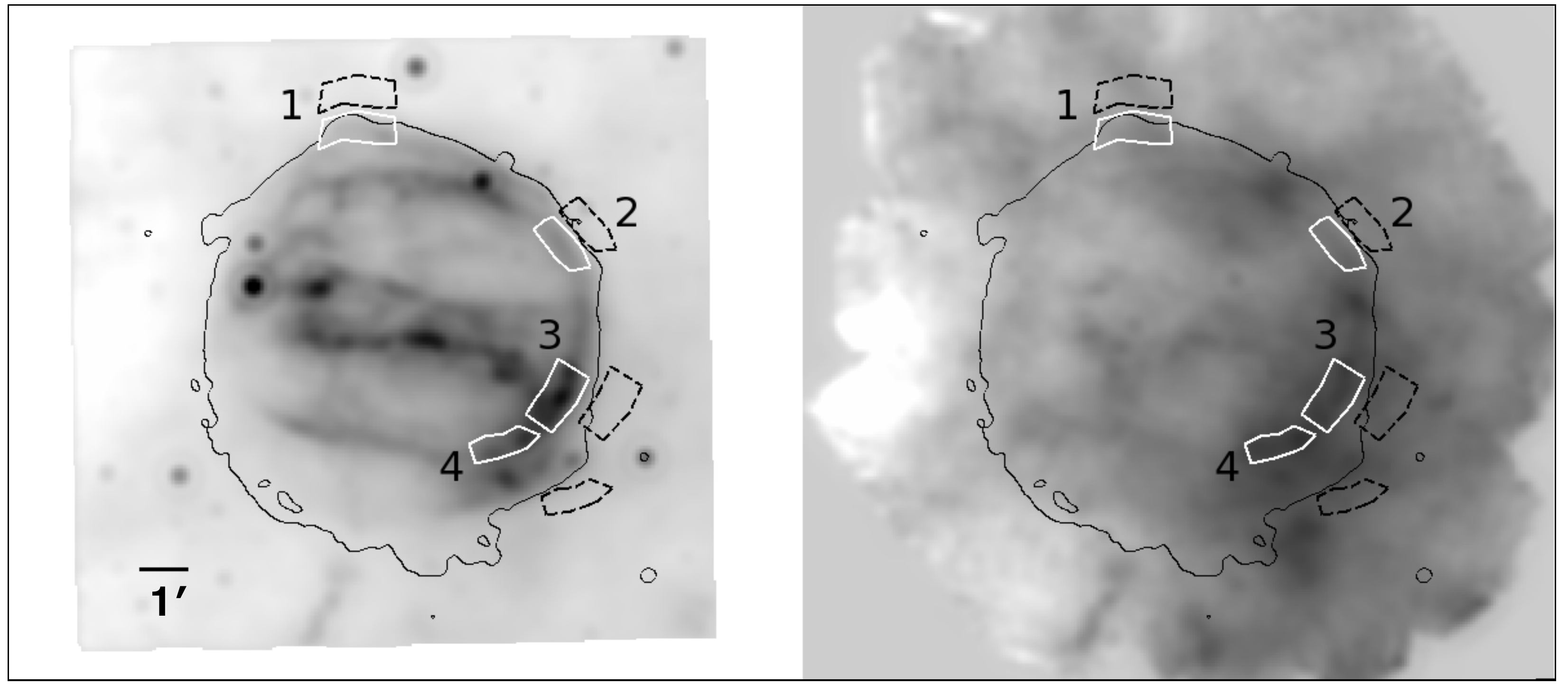}  
   \caption{ Left: {\it Spitzer} MIPS Image of \g292\, at 24  $\micron$  (Ghavamian \etal\, 2012).       Right: The {\it Herschel} PACS 100 $\micron$ image of \g292.  The black
   contour marks the outer edge of the supernova remnant as measured from the \chan\, X-ray image of the SNR (Park \etal\, 2007).  The
   spatial resolution and pixel scales of both images shown here have been coarsened to that of the 70 $\micron$ MIPS band of {\spitzer} (see text for details).  The integrated fluxes
   from \g292\, at 24 $\micron$ and 100 $\micron$ reported in Figure~6 were obtained from these images.  Regions used for extracting the IRS spectra in Figure~8 are
     marked with the white regions, and the corresponding backgrounds are marked with dotted black regions.
  }
   \label{fig:fig5}
\end{figure}

\begin{figure}
\centering
\includegraphics[width=7in]{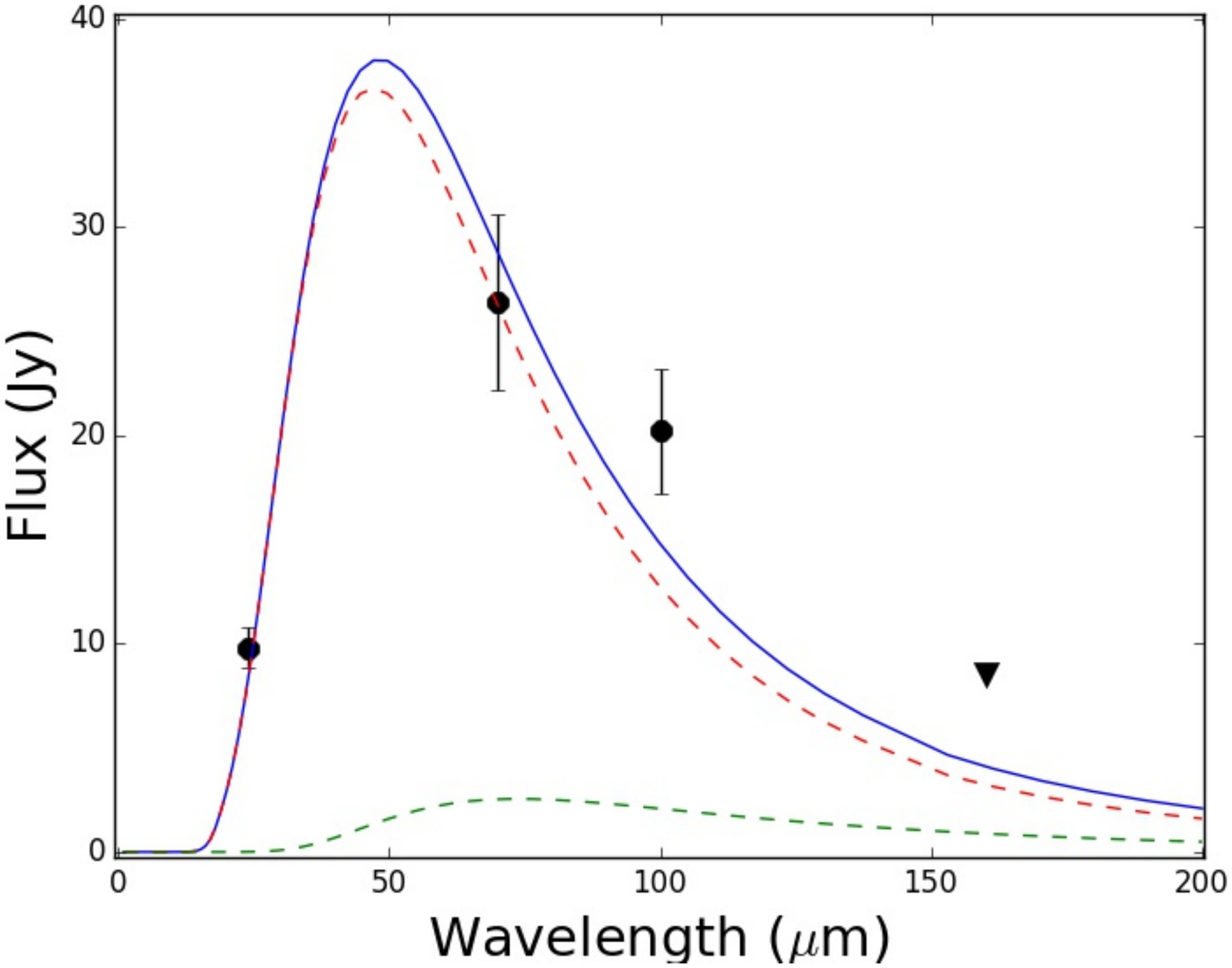}  
   \caption{ Global infrared SED of \g292, with data points taken
     from the MIPS 24 $\micron$ and 70 $\micron$ images (Ghavamian
     \etal\, 2012) as well as the PACS 100 $\micron$ image and PACS
     160 $\micron$ image (the latter is an upper limit, as no obvious
     emission is detected from \g292\, at 160 $\micron$).  
     A best-fit model to the 24, 70 and 100 $\micron$ data points is indicated with the blue line,
     with individual components of silicate grains indicated with the red dashed line and graphite grains 
     indicated with the green dashed line (see text for details).     
      }
   \label{fig:fig6}
\end{figure}

\begin{figure}
\centering
\includegraphics[width=6in]{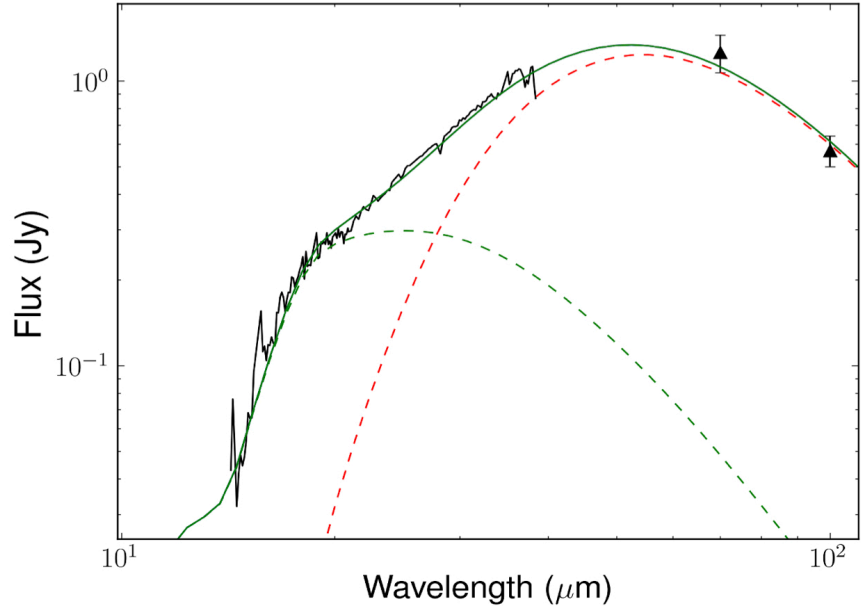}  
   \caption{ Integrated spectrum of the radiatively shocked O-rich
     Spur.  The 15-40 $\micron$ component was obtained from the
     \spitzer\, IRS spectral map of G292.0+1.8 (Ghavamian \etal\, 2012),
     while the data points at 70 $\micron$ and 100 $\micron$ are from
     \spitzer\, MIPS and Herschel PACS, respectively. The spectrum
     reflects contributions from both shocked circumstellar dust and
     ejecta dust, due to line of sight superposition of the two
     components.  The spectrum is fit with a hot component (green dotted line) and
     cool component (red dotted line). }
   \label{fig:fig7}
\end{figure}

\begin{figure}
\centering
\includegraphics[width=6in]{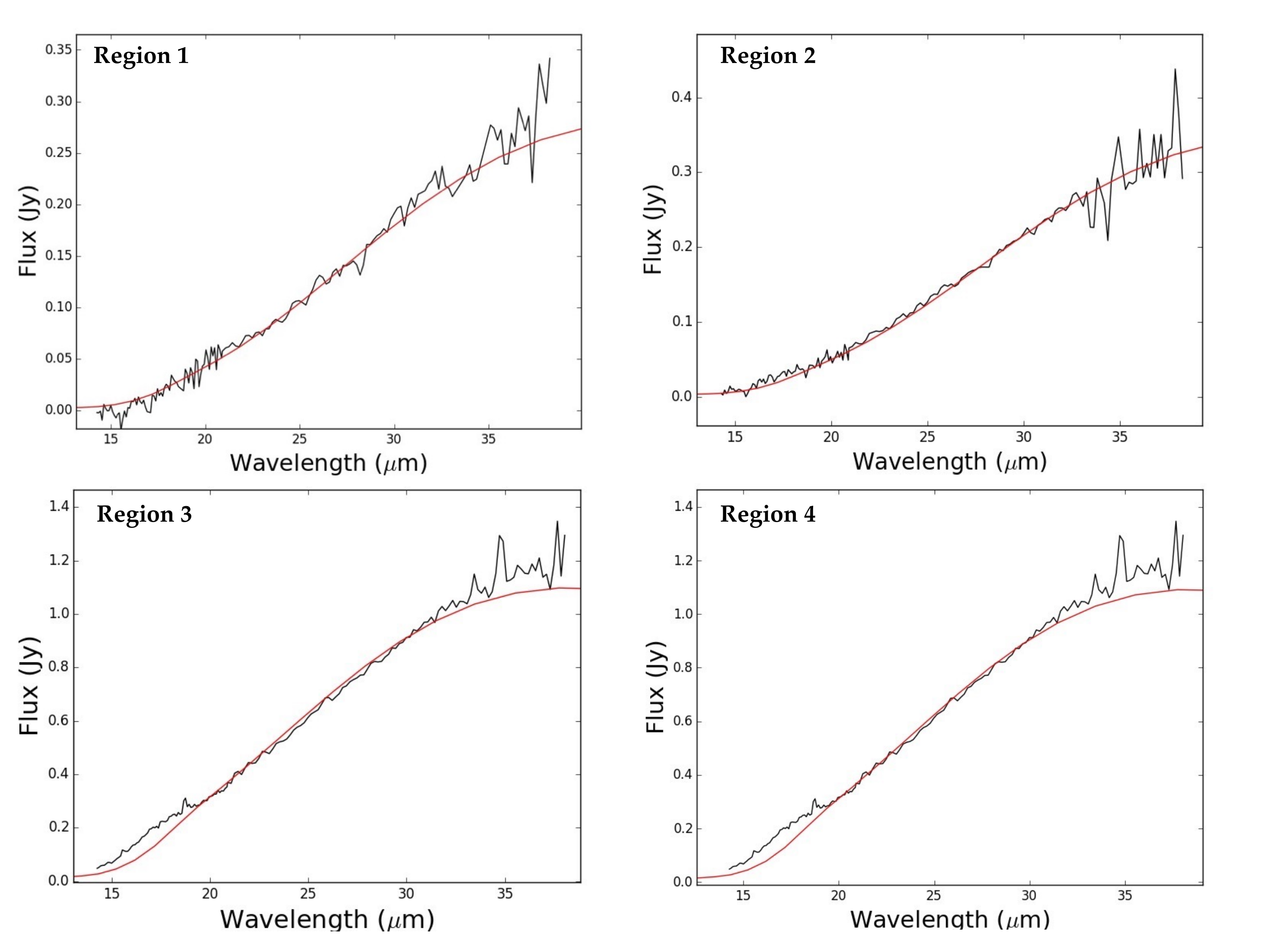}  
   \caption{ Background-subtracted \spitzer\, IRS spectra of four locations around the rim of
     \g292\, (regions marked in Figure \ref{fig:fig5}).  Shock model fits
     to each spectrum are marked with red lines.  }
   \label{fig:fig8}
\end{figure}

\end{document}